\documentstyle[12pt]{article}
\global\arraycolsep=2pt
\newcommand{\ra}{\rangle}
\newcommand{\la}{\langle}
\begin{document}

\begin{titlepage}
\begin{flushright}
SMU-HEP-93-24\\
hep-ph/9311377\\
November 1993
\end{flushright}

\vspace{0.3cm}

\begin{center}
\Large\bf Singular defects with fractional topological charge\\
in $\sigma$  models and gauge theories.\footnote
{{\it Contribution to the Proceedings of
 Lattice-93, Dallas, 12-16 October, 1993;}}

\end{center}
\vspace{0.8cm}

\begin{center}

{\bf Ariel R.Zhitnitsky}
\footnote
{On leave of absence from Budker Institute of Nuclear
Physics, Novosibirsk 630090, Russia.
e-mail addresses: arz@smuphy.physics.smu.edu ~~~ariel@sscvx1.ssc.gov .
}\\
\vspace{0.3cm}
{\it Physics Department, SMU,
Dallas, Texas, 75275-0175.}

\end{center}

\begin{abstract}
A novel class of self-dual solutions in $\sigma$ models and gauge theories
is considered. The contribution of the corresponding fluctuations to the chiral
condensate is calculated. We discuss the few tightly connected problems,
such as the $U(1)$ problem, the $\theta$ dependence and
the chiral symmetry breaking
 within a framework of this approach. Arguments in favour of
significance of the configurations with fractional
topological charge are given.
\end{abstract}
\end{titlepage}
 \section{Motivation. Why fractional topological
charge should be considered?}

Let me start these notes from the well known results concerning the
chiral condensate in supersymmetric models.
 More specifically, let us consider 2 dimensional SUSY $CP^{N-1}$ model,
\cite{Wit2}. As is known the model possesses naive
$U(1)$ chiral symmetry, which is broken by anomaly. However, the
discrete symmetry $\sim Z_N$ is conserved. At large $N$ this model
can be explicitly solved and shows up the nonzero value for
$\la\bar{\psi}\psi\ra$, which corresponds to discrete symmetry breaking
phenomenon.

At the same time, the instanton can ensure non-zero value
for the correlator  $\la\prod^{N}_{i=1}\bar{\psi}\psi(x_i)\ra$
 only, and not for
the condensate $\la\bar{\psi}\psi\ra$.
 The reason is trivial and related to the fact,
that one instanton transition is always accompanied by the emission
of $2N$ fermionic zero modes. By {\it clustering}, at $(x_i-x_j)
\rightarrow\infty$ this relation implies a nonvanishing
magnitude for condensate as well, however $\la\bar{\psi}\psi\ra_{inst}=0$
because we have $2N$ and not $2$ zero modes. Besides that,
in according with Witten index \cite{Wit1} we have $N$ different
vacuum states classifying by the phase of condensate
$\la\bar{\psi}\psi\ra\sim\exp(\frac{2\pi ik +\theta}{N})$. Let us note
that $\theta$ dependence comes through $\frac{\theta}{N}$. Such
a function can be periodic in $\theta$ with period $2\pi$ only if there are
many ($N$) vacuum states for given values of $\theta$.
I have to note that the same situation takes place in the supersymmetric
gluodynamics as well as   in SQCD. I refer to the review
paper \cite{Amat} on this subject in supersymmetric
four dimensional  models, but here
I want to  make a remark that such behavior is not specific for
SUSY models.

In particular,
the analogous $\theta/N$
dependence was discovered in gluodynamics at large N
\cite{Wit},\cite{Ven},
\cite{Wit3}. In these papers was argued that the vacuum
energy at large N appears in the form $E\sim E(\theta/N)$.
  This fact actually
is coded in the effective lagrangian containing the
multi-branched logarithm $\log\det (U)$.
In the Veneziano approach \cite{Ven} the same fact can be seen
from the formula for multiple derivation of  the topological
density $Q $
 with respect to $\theta$ at $\theta =0$.
\begin{equation}
\label{a}
\frac{\partial^{2n-1}}{\partial\theta^{2n-1}}\la Q(x)\ra\sim
(\frac{1}{N})^{2n-1} \ \ ,n=1,2...
\end{equation}
 It is clear that it corresponds to the following $\theta$ dependence
of the topological density $\la Q\ra\sim\sin\frac{\theta}{N}$.
Let me remind, that the $\theta$ is the physical
 parameter of the theory and
the $\theta$-
dependence of physics is linked to the $U(1)$ problem \cite{Wit},\cite{Ven}.
Indeed,
  if we believe
that the resolution of the $U(1)$ problem appears within the framework
of these papers, we must assume that the
correlator
\begin{equation}
\label{1}
K=i\int d^4x\la 0|T{Q(x),Q(0)}|0 \ra
\end{equation}
is nonzero in pure gluodynamics. But the topological
susceptibility $K$ is nothing but divergence of $\la Q\ra $ with
respect to $\theta$. As is known $K\sim\frac{1}{N}$. It demonstrates
one more times  that $\theta$ parameters comes to the theory through
$\theta/N$.

The question we want to raise can be formulated as follows.
How can one reproduce $\theta/N$ dependence in the theory with
integer topological charges only?  Our answer is:
The configurations with {\it fractional topological charges
with finite action should be introduced to the theory}.
\section{Basic assumptions}
$\bullet$ I extend the class of admissible gauge transformation in
gluodynamics. Thus, I allow the configurations with
fractional topological charge (one half for $SU(2)$ group)
in the definition of the functional integral.
I call this configurations {\it toron}\footnote{ We keep the
term "toron", introduced in ref.\cite{Hoo1}. By this means
we emphasize the fact that the considering solution
minimizes the action and carries the topological charge
$Q=1/2$,i.e. it possesses all the characteristics ascribed
to the standard toron \cite {Hoo1}.}.
This extension means that
a multivalued functions will appear in the functional
integral. However, the main physical requirement is: {\it all
gauge invariant values must be singlevalued}. Thus, the
different cuts accompany the multivalued functions should
be unobservable. Let us note that at large distances the
{\it toron looks like a singular
gauge transformation }.

$\bullet$ The next main point of the toron approach may be
formulated as follows. We hope that in the functional
integral of the gluodynamics,  only certain field configurations  (
the toron of all types) are important.
In this case, the consistency of these assumptions can be checked
 by
considering a few simple models, where, on the one hand,
the results are well known beforehand and, on the other
hand they can be reproduced by the toron calculations
\cite{Zhi1}.
\section{  $0(3)\sigma$ model. Lessons and Experience.}
We define the action and the topological charge of the
supersymmetric
$0(3)\sigma $model, equivalent to the $CP^1$ theory as follows
\cite{Wit2}:
\begin{equation}
\label{3}
S=\frac{1}{f}\int d^2x|D_{\mu}n|^2, ~~
 D_{\mu}=\partial_{\mu}-iA_{\mu},
\end{equation}
\begin{equation}
 F_{\mu\nu}=\partial_{\mu}A_{\nu}-\partial_{\nu}A_{\mu},~~
 Q=\frac{1}{4\pi}\int d^2x \epsilon_{\mu\nu}F_{\mu\nu},
\end{equation}
Here $n_{\alpha}$ is a complex 2-component unit spinor, transforming
 according to the fundamental representation of $SU(2)$, $A_{\mu}
=-i\bar{n}\partial_{\mu}n$
is an auxiliary gauge field, and we have shown only bosonic part
of the action.  The classical solution in this language is determined
by analytical function $P_{\alpha}(z)$, where $z=x_1+ix_2$
and $n_{\alpha}= P_{\alpha}/|P|$.
In particular, the standard instanton solution takes the form:
\begin{equation}
\label{4}
 n_{inst}=\frac{1}{\sqrt{|z-z_0|^2+\rho^2}}\left(^{~~\rho} _{z-z_0} \right)
_{z\rightarrow\infty}\Rightarrow e^{i\phi}\left(^0 _1 \right)
\end{equation}
and becoming a pure gauge transformation at large distances.
We wish to describe the solution $n_{t}$ which at large distances
looks as a pure gauge field
$n_{t}(z\rightarrow\infty)\Rightarrow e^{i\phi/2}\left(^0 _1 \right)$
with one half phase $\phi/2$ instead of integer phase $\phi$
(\ref{4}) in order
to describe one half topological charge.
Besides that, we would like to regularize this behavior at
small distances by parameter $\Delta\rightarrow 0$ in the
 very special way in order to preserve the selfduality equation:
\begin{equation}
\label{5}
n_{t}=\frac{1}{\sqrt{|z-z_0|+|\Delta|}}\left(^{~~
\sqrt{|\Delta|}} _{\sqrt{z-z_0}} \right)
_{z\rightarrow\infty}\Rightarrow e^{i\phi/2}\left(^0 _1 \right)
\end{equation}
As was expected, this solution is a double-valued function and it is defined
on a covering space. Main physical requirement to these configurations
is: all {\it gauge invariant values should be singlevalued}.
In particular, the classical action is $ S_{t}=1/2S_{inst}$
 and the corresponding density
is  singlevalued and in the limit
$\Delta\rightarrow 0$ goes to $\delta^2(x-x_0)$ function:
\begin{equation}
\label{6}
S= \lim_{\Delta\rightarrow 0}\frac{\Delta}{2f}\int\frac{d^2x}
{|x|(|x|+|\Delta|)^2}\rightarrow \frac{\pi}{f}\int d^2x\delta^2(x)
\end{equation}
With regularized expression (\ref{5}) it can be easily checked that
the number of fermionic (and bozonic) zero modes (ZM) equals two and
not four, as in instanton case.
The toron measure in the supersymmetric version of the model
is given by
\begin{equation}
\label{7}
Z_{t}\sim M_0^2 d^2z_0\frac{d^2\epsilon}{M_0}e^{\frac{-\pi}{f}}=
md^2z_0d^2\epsilon,~m=M_0e^{\frac{-\pi}{f(M_0)}}
\end{equation}
where $m$ is renormalization invariant combination.
Now we are ready to calculate the chiral condensate in the model.
Substituting the ZM in place of $\psi$, and recalling the
integration
over the collective variables satisfies
$\int\epsilon^2d^2\epsilon=1$, we verify that
\begin{equation}
\label{8}
\la Q_5=2|\bar{\psi}_L {\psi}_R|Q_5=0\ra\sim\int d^2z_0
\bar{\psi}_0  {\psi}_0 =m
\end{equation}
Because the transition amplitude (\ref{8}) is non-zero, and because the toron
transition changes the chiral charge $Q_5$ by two units, the true
physical states$|\Omega_{\pm}\ra$ must be superposition of the states
$|Q_5=0,2\ra$. These two
vacuum states are true physical vacua of spontaneously broken
discrete chiral symmetry:
\begin{equation}
\label{9}
\la\Omega_{\pm}|\bar{\psi}_L {\psi}_R|\Omega_{\pm}\ra= \pm m
\end{equation}
Besides that, it can be  explicitly checked that these physical
vacua provide the correct $\theta/2$ dependence and thus,
the $\theta$ evolution from $\theta=0$ to $\theta=2\pi$ renumbers
two degenerate states$\Omega_{\pm}$.
Let me repeat, that as soon as we allowed one half topological charge,
{\it the number of the classical vacuum states is increased} by the
same factor two in comparison with the standard classification, counting
only integer winding numbers $|n\ra$. This result is in a full
agreement with  large $N$ results  presented above.
Analogous calculations can be done in four-dimensional case   and
we refer to the original papers \cite{Zhi1}.
\section{Conclusion, Interpretation, Problems.}
$\bullet$
We interpret the standard instanton as the pseudoparticle,
 constructed from
these singular points defects. In particular, for 2-dimensional
$CP^{N-1}$ model we interpret $2N$ boson ZM accompanied by instanton
as translation modes for $N$ different torons.
Four dimensional instanton for
any gauge group $G$ can be interpreted in the same way.
In this case as is known the number of bosonic ZM equals
$4C(G)$, where $C(G)$ is Casimir operator ($C(SU(N))=N$).
This number we interpret as translations of $C(G)$ torons.
This conjecture, in particular, is in agreement with formula
for the instanton
measure in supersymmetric gluodynamics
\begin{equation}
\label{10}
Z_{inst}\sim \prod_{i=1}^{C(G)}d^4x_id^2\epsilon_{i}
(e^{-\frac{8\pi^2}{g^2C(G)}})^{C(G)}\sim \prod_{i}^{C(G)}Z_{t}(i),
\end{equation}
ensures the correct renormalization invariant dependence and
gives the correct dependence on $\theta$ for gluino condensate.

$\bullet$
The direct consequence of   our definition of the
functional integral is the appearing of the new quantum
number classifying the vacuum states. Indeed, as soon as we
allowed one half topological charge, the number of the
classical vacuum states is increased by the same factor two
in comparison with a standard classification, counting only
integer winding numbers $|n \ra$. This is exactly what
we observed from the large $N$ analysis.

Of course, vacuum transitions eliminate this degeneracy.
However the trace of enlargement number of the classical
vacuum states does not disappear. Vacuum states now
classified by two numbers : $0\leq \theta <2\pi$ and
$k=0,1$.  These is in agreement with large $N$ results
where the nontrivial $\theta$ dependence in pure YM theory
comes through $\theta/N$ \cite{Wit3} at large $N$ and we
had $N$ additional states for each given $\theta$.

$\bullet$
 I would like to stress that a lot of
problems ( like the $\theta$ dependence, the $U(1)$
problem, the counting of the discrete number of vacuum
states,  the nonzero value of the vacuum
energy and so on...) can be described in a very simple
manner from this uniform point of view.

$\bullet$
Main question to Lattice Community:
How can one describe these singular self-dual (with finite action)
defects on the Lattice?
Some results on this subject have been presented by
Antonio Gonzalez-Arroyo to this Proceedings.
This work is supported by the TNRLC grant No 528428

\end{document}